\documentclass[onecolumn,showpacs,preprintnumbers,amsmath,amssymb]{revtex4}

\usepackage{graphicx}
\usepackage{dcolumn}
\usepackage{bm}

\begin{document}

\preprint{APS/123-QED}

\title{Using cosmological perturbation theory to distinguish between General Relativity and Unimodular Gravity}

\author{Marcelo H. Alvarenga}
 \altaffiliation{}
 \email{marcelo.alvarenga@edu.ufes.br}
\affiliation{Núcleo Cosmo-ufes \& Departamento de Física, UFES, Vitória, ES, Brazil
}%

\author{Júlio C. Fabris}
 \altaffiliation{}
 \email{julio.fabris@cosmo-ufes.org}
\affiliation{Núcleo Cosmo-ufes \& Departamento de Física, UFES, Vitória, ES, Brazil
\\
National Research Nuclear University MEPhI, Kashirskoe sh. 31, Moscow 115409, Russia
}%

\author{Hermano Velten}%
\email{hermano.velten@ufop.edu.br}
\affiliation{%
Departamento de F\'isica, Universidade Federal de Ouro Preto (UFOP), Campus Universit\'ario Morro do Cruzeiro, 35.400-000, Ouro Preto, Brazil}

\date{\today}

\begin{abstract}
Unimodular Gravity is one of the oldest geometric gravity theory alternative to General Relativity. Essentially, it is based 
on the Einstein-Hilbert Lagrangian with an additional constraint on the determinant of the metric. It can be explicitly shown that Unimodular Gravity can be recast as General Relativity in presence of a cosmological constant.
This fact has led to many discussions on the equivalence of both theories at classical and quantum levels. Here we present an analysis focused on the classical scalar perturbations around a cosmological background. The discussion is extended for the case where a non-minimal coupled scalar field is introduced. Our results indicate that the equivalence is not verified completely at perturbative level.\end{abstract}

\pacs{04.20.Cv, 04.60.-m, 98.80.-k}
\maketitle

\section{Introduction}

General Relativity (GR) theory aims to describe gravity as a manifestation of the geometry of the space-time in four dimensions. It is based on the mathematics of differential geometry. The theory is invariant by the full group of diffeomorphism transformations. It has led to important previsions like the existence of black holes and gravitational waves. At same time,
it admits very successful applications for the description of the universe as a whole, leading to the standard cosmological model. The latter explains in a quite simple way all observational data, even at the price of introducing a dark sector composed of two exotic components (dark matter and dark energy), not detected directly until now, and also a primordial inflationary phase driven by a new field, the inflation, whose nature is still a matter of debate. On the other hand, GR is plagued by the presence of singularities and, at the same time, it has not been quantized until now in a fully consistent way.
These last problems led to an intensive search for an alternative description of the gravitational phenomena, in general keeping the geometrical approach, which may cope with the existence of the dark sector, the inflationary phase, being free of singularities and admitting a consistent quantum version.

GR has been proposed in 1915 and soon after alternative formulations appeared. In 1919, the unimodular constrained version of GR has been formulated, leading to what is now known as Unimodular Gravity (UG) \cite{ein,pauli}. In UG the determinant of the metric is fixed as a constant, in occurrence, equal to $1$. Whereas this can be viewed as a choice of the coordinate system, it has some important consequences. First of all, UG is invariant by a subclass of transformations, called transverse diffeomorphism (TD) \cite{trans}. Moreover, the field equations are traceless implying the absence of information about geometrical quantities like e.g., the Ricci scalar $R$. Matter can be coupled in a particular way to the geometrical sector, preserving the traceless character of the equations. If the conservation of the canonical energy-momentum tensor is imposed, the UG field equations imply the GR equations in presence of a cosmological constant $\Lambda$, which appears as an integration constant. This is generally viewed as an advantage with respect to GR. Moreover, it is generally argued that UG has some improvements over GR at quantum level. Many discussions exist, on the other hand, on the possible equivalence between GR e UG, but the  restriction of the invariance of UG to the TD group seems to indicate that this equivalence is not complete, even if it may appear in some contexts (see, for example, Refs. \cite{wei-ug, ant,esp} and references therein). 

One of the possible reasons for the non-equivalence between GR and UG comes from the conservation of the canonical energy-momentum tensor. In GR the conservation laws are direct consequences of the invariance with respect to general diffeomorphism transformation \cite{wald}. This property is reflected in the use of the Bianchi identities. For UG the invariance with respect to the TD does not allow to obtain the same conservation laws as in GR. Moreover, the UG field equations are traceless and the application of the Bianchi identities leads to a relation where the divergence of the canonical energy-momentum tensor
is not necessarily zero. However, we can impose by hand the same conservation laws as in GR. In doing so, UG leads to the GR field equations in presence of a integration constant (identified as the cosmological constant) is obtained. Does this means an equivalence between GR and UG? First we must remember that the usual conservations constitute a choice in UG. Moreover, in UG, even with this choice, the invariance remains dictated by the restricted TD, instead of the full diffeomorphism.

In this text, we will discuss the issue of the perturbative features of UG in comparison to GR. It will be shown that, while for the vacuum case there seems to be an equivalence, the presence of matter may modify this conclusion. This perturbative non-equivalence can be extended even for the vacuum solutions when an extension of the unimodular structure to non-minimal coupled scalar field is implemented. In next section, we settle out both the fundamental equations for the cosmological background both for GR and UG. We turn to the the specific cosmological background solutions in section III. In section IV the perturbative issue is discussed. It is shown the equivalence of the results in both context in the vacuum case, and it is discussed the case
when matter is present, for which the equivalence may not be verified anymore. The unimodular version of the Brans-Dicke theory is discussed in section V. In section VI we present our final considerations.

\section{Background cosmological structure}
 
 The GR equations in the presence of a cosmological constant and of a matter sector, as deduced from the Einstein-Hilbert Lagrangian, are,
 \begin{eqnarray}
 \label{grfe1}
 R_{\mu\nu} - \frac{1}{2}g_{\mu\nu}R = 8\pi G T_{\mu\nu} + g_{\mu\nu}\Lambda. 
 \end{eqnarray}
 The application of the Bianchi identities leads to the energy-momentum tensor $T^{\mu\nu}$ conservation:
 \begin{eqnarray}
 \label{grfe2}
 {T^{\mu\nu}}_{;\mu}  = 0.
 \end{eqnarray}
The conservation laws related to the energy-momentum tensor can be alternatively  deduced from the invariance of the Einstein-Hilbert Lagrangian by diffeomorphic transformations \cite{wald}.

The UG equations can also be deduced from the Einstein-Hilbert Lagrangian but through the introduction of a constraint
on the determinant of the metric. For details, see \cite{fabris1,fabris2}.
With this procedure, the UG field equations read,
\begin{eqnarray}
\label{ugfe1}
 R_{\mu\nu} - \frac{1}{4}g_{\mu\nu}R = 8\pi G\biggr\{T_{\mu\nu} - \frac{1}{4}g_{\mu\nu}T\biggl\}.
 \end{eqnarray}
 Equations (\ref{ugfe1}) are valid even if we introduce a cosmological constant in the action, i.e., after solving the constraints represented by the Lagrangian multiplier, the cosmological term disappears from the final equations.
 The application of the Bianchi identities leads to,
 \begin{eqnarray}
 \label{ugfe2}
 \frac{R^{;\nu}}{4} = 8\pi G\biggr\{{T^{\mu\nu}}_{;\mu} - \frac{T^{;\nu}}{4}\biggl\}.
 \end{eqnarray}
 
 Equation (\ref{ugfe2}) deserves some comments. In opposition to what happens in GR, the TD, on which is based UG, does not lead necessarily to the conservation equations (\ref{grfe2}). Instead, it predicts a non-vanishing divergence of the energy-momentum tensor given by,
 \begin{eqnarray}\label{theta}
 {T^{\mu\nu}}_{;\mu}  = \Theta^{;\nu},
 \end{eqnarray}
 $\Theta$ being, in principle, an unknown scalar function. Nothing forbids to impose the extra condition given by (\ref{grfe2}). If this extra condition is imposed, (\ref{ugfe2}) can be integrated leading to (\ref{grfe1}). However, in doing so,
 $\Lambda$ appears as an integration constant. This is the basis of the usual remark that UG may alleviate the cosmological constant problem, since $\Lambda$ now is not associated with the vacuum energy or to a geometric zero order Lovelock invariant.
 
 In fact, if the conservation of the energy-momentum tensor is imposed, by vanishing the right hand side of  (\ref{theta}), then Eq. (\ref{ugfe2}) implies,
  \begin{eqnarray}
 R^{;\nu} = - 8\pi G T^{;\nu} \quad \Rightarrow  \quad R + 8\pi G T = \mbox{constant}.
 \end{eqnarray}
 Identifying the constant in the right hand side of the above result with $- 4 \Lambda$, the GR equations with a cosmological term are recovered.
 
 Of course, we can also keep (\ref{ugfe2}), since it results from the application of the Bianchi identities via the identification
 \begin{eqnarray}
 \label{ce1}
\Theta = \frac{R}{4} +  2\pi G T.
 \end{eqnarray}
 But, again, this is a choice and many other possible ones are admitted since $\Theta$ is not determined from the unimodular construction.
 
 Let us for the moment make the choice (\ref{ce1}) in order to verify the consequences in a specific cosmological context.
 In any case, the choice $\Theta = 0$ implies GR with a cosmological constant, with the remaining issue of the interpretation of this constant.
 
 For sake of simplicity, let us consider the flat FLRW metric given by,
 \begin{eqnarray}
 \label{flrw}
 ds^2 = dt^2 - a^2(t)(dx^2 + dy^2 + dz^2).
 \end{eqnarray}
 In our approach to the UG, the unimodular condition is $\sqrt{-g} = \xi$ where $\xi$ is an arbitrary reference (external, if one prefers) tensorial density, allowing to choose freely the coordinate system, in opposition to the original choice $\sqrt{-g} = 1$, which fixes the coordinate system.
 
 Applying the metric (\ref{flrw}) on the GR equations we obtain the following equations of motion:
 \begin{eqnarray}
 H^2 &=& \frac{8\pi G}{3} \rho + \frac{\Lambda}{3},\\
 \dot\rho + 3 H(\rho + p) &=& 0,
 \end{eqnarray}
 with $H = \dot a/a$, the Hubble function. We must specify an equation of state connecting $\rho$ and $p$. In doing so, we are left with two equations for two variables, $\rho$ and $a$.
 
 On the other hand, the same metric applied to (\ref{ugfe1},\ref{ugfe2}) implies,
 \begin{eqnarray}
\label{em1}
\dot H &=& - 4\pi G(\rho + p),\\
\label{em2}
\ddot H + 4H\dot H &=& - 4\pi G[\dot\rho + \dot p + 4H(\rho + p)]. 
\end{eqnarray}
The main point here is that, inserting (\ref{em1}) into (\ref{em2}) we obtain an identity. Hence, both equations have the same content: the system is underdetermined, having more unknown functions than equations. This is a consequence of the fact that the UG equations (\ref{ugfe1}) are traceless, leading to no information on the Ricci scalar, in opposition to the GR equations.
Moreover, equation (\ref{em1}), as (\ref{em2}), is sensitive only to the combination $\rho + p$ which is the enthalpy density of the system.
In Ref. \cite{fabris1} it has been made the choice that matter behaves as radiation, as suggested by the traceless character of the field equations. This choice has led to many drastic implications at perturbative level, as discussed in Ref. \cite{fabris1}. In particular, as we will discuss later, there is a direct transition from the radiation era to the de Sitter era.
However, due to the structure of UG, fluctuations grow strongly even during the radiation phase, what may assure the formation of structures in the universe.

In what follows, we will focus first on the vacuum case. This will allow us to verify, for a specific configuration, to which extent the UG is, at least classically, equivalent to GR as frequently evoked.

\section{UG and GR vacuum cosmological solutions}

In the vacuum case, $\rho = p = 0$. Notice however that $\Lambda$ may remain different from zero in GR, leading to,
\begin{eqnarray}
H^2 = \frac{\Lambda}{3}.
\end{eqnarray}
The first remark is that only $\Lambda \geq 0$ is possible, that is, a de Sitter or Minkowski solution, excluding the Anti-de Sitter case ($\Lambda < 0$). The Hubble function is constant, implying, for $\Lambda > 0$, an exponential solution for the scale factor.
Using the definition of the Ricci scalar,
\begin{eqnarray}
R = - 6\biggr(\dot H + 2H^2\biggl),
\end{eqnarray}
the solution implies $R$ constant and negative, in agreement with a maximally symmetric de Sitter space-time.
The solution for the scale factor is given by,
\begin{eqnarray}
a \propto e^{\pm Ht},
\end{eqnarray}
describing either an exponentially expanding or a contracting universe. For the applications for the inflationary universe, only the expanding solution is used. Remark that the Minkowski case corresponds to $\Lambda = 0$. Hence, the Minkowski and the de Sitter are the possible space-times in the vacuum cosmological solutions of the GR theory in presence of a cosmological constant, Minkowski being the trivial one.

For UG, the vacuum case leads to two simple equations:
\begin{eqnarray}
\label{ugcs1}
\dot H &=& 0, \\
\label{ugcs2}
\dot R &=& 0.
\end{eqnarray}
The last one implies $R = - 2\Lambda_U$, where now $\Lambda_U$ is an integration constant which be related to the cosmological constant in the GR context. The factor $-2$ has been introduced just to make this connection easier.
Again, the solution of (\ref{ugcs2}) implies,
\begin{eqnarray}
H = \pm \sqrt{\frac{\Lambda_U}{3}} \quad \rightarrow \quad a \propto e^{\pm \sqrt{\frac{\Lambda_U}{3}} }.
\end{eqnarray}
The constant $\Lambda_U$ must be positive or zero, the last case leading to the trivial Minkowski space-time. The Ricci scalar leads to the equation,
\begin{eqnarray}
H^2 = \frac{\Lambda_U}{3},
\end{eqnarray}
that is the Friedmann equation in GR for the vacuum case in presence of a cosmological constant. But the origin of this {\it cosmological constant} is quite different: it does not come from the Lagrangian with a cosmological term, but as an integration constant. In general it is argued that this (important) formal difference between the Friedmann equation in GR and UG alleviates the cosmological constant problem.

What happens if matter is introduced? The main point is that, as it was already briefly discussed above, it is impossible to solve the UG cosmological equations in this case since there is just equation (\ref{em1}) for two variables $\rho$ and $a$: the equation can not be solved without an extra assumption. If the conservation of the energy-momentum tensor is imposed, we obtain the same solutions as GR in presence of a cosmological constant, the cosmological term appearing, as in the vacuum case, as an integration constant. Another path has been followed in Ref. \cite{fabris1}, exploring the traceless character of the field equations in UG, which implies that the matter sector must be also traceless indicating that a radiative fluid is the natural choice for the matter sector. In this case, the usual radiative solutions in presence of a cosmological constant are recovered. However, the perturbative behavior strongly differs from the GR case.

Now, we turn to the perturbative analysis of the vacuum solution, introducing some comments for the case matter is present. 

\section{Perturbative analysis}

At perturbative level the differences between GR an UG becomes more evident. First of all, while in GR the general diffeomorphism invariance allow to fix a coordinate condition or to use a gauge invariant set of variables when the perturbative analysis is made, in UG the choices are much more restricted due to the invariance by the TD. This can been seen by considering the unimodular constraint,
\begin{eqnarray}
\sqrt{-g} = \xi.
\end{eqnarray}
As already discussed, $\xi$ is a fixed external quantity. Perturbing the metric, 
\begin{eqnarray}
\tilde g_{\mu\nu} = g_{\mu\nu} + h_{\mu\nu},
\end{eqnarray}
and preserving the unimodular constraint, we are led to the relation,
\begin{eqnarray}
h = h^\rho_\rho = 0.
\end{eqnarray}

Let us now consider the general perturbed metric restricted to the scalar sector:
\begin{eqnarray}
ds^2 = a^2\biggr\{(1 + 2\phi)d\eta^2 - 2B_{,i}dx^i d\eta - \biggr[(1 - 2\psi)\delta_{ij} + 2E_{,i,j}\biggl]dx^idx^j \biggl\}.
\end{eqnarray}
Here on we follow closely the notation of Ref. \cite{bmf}. The condition $h = 0$, implies,
\begin{eqnarray}
\label{pc1}
\phi - 3\psi - \nabla^2E = 0.
\end{eqnarray}
The newtonian gauge is obtained by fixing $B = E = 0$, implying $\phi = 3\psi$. This condition contradicts the other condition obtained from the perturbed equations when anisotropic pressure is absent, $\phi = \psi$, leading to $\phi = \phi = 0$ and no perturbation is present. The situation with the synchronous coordinate condition is more involved, since this condition implies $\phi = B = 0$, leading to $\nabla^2 E = - 3\psi$, which can be re-expressed as $h_{kk} = 0$. However, if the conservation of the energy-momentum is imposed, $h_{kk}$ is directly related to the matter perturbation: $h_{kk}$ being zero, there is no matter perturbation also. The situation changes when the conservation of the energy-momentum tensor is not imposed, as we will se latter. The gauge invariant formalism \cite{bmf} can always be used, but with the additional condition (\ref{pc1}).

\subsection{The perturbed equations in the gauge invariant formalism}

The perturbed field equations in GR using the gauge invariant formalism, with the hydrodynamical approach, read \cite{bmf}:
\begin{eqnarray}
\label{ge1}
- 3{\cal H}(\Psi' + {\cal H}\Phi) + \nabla^2\Psi &=& 4\pi Ga^2\delta\bar\rho,\\
\label{ge2}
\biggl\{\Psi' + {\cal H}\Phi\biggl\}_{,i} &=& - 4\pi G(\rho + p)a^3\delta\bar u ^i,\\
\label{ge3}
\biggr[\Psi'' + {\cal H}(2\Psi' + \Phi') + (2{\cal H}' + {\cal H}^2)\Psi + \frac{1}{2}D\biggl]\delta_{ij} - \frac{1}{2}D_{,i,j} &=& 
4\pi Ga^2\delta\bar p\delta_{ij},
\end{eqnarray}
where $D = \Phi - \Psi$. Moreover, ${\cal H} = a'/a$, the primes indicating derivative with respect to the conformal time.

No anisotropic pressure is considered. The non-diagonal terms of (\ref{ge3}) $i \neq j$ lead to $D = 0$, implying $\Phi = \Psi$.
The resulting equations are,
\begin{eqnarray}
\label{ge1-bis}
- 3{\cal H}(\Phi' + {\cal H}\Phi) + \nabla^2\Phi &=& 4\pi Ga^2\delta\bar\rho,\\
\label{ge2-bis}
\biggl\{\Phi' + {\cal H}\Phi\biggl\}_{,i} &=& 4\pi G(\rho + p)a^3\delta\bar u ^i,\\
\label{ge3-bis}
\Phi'' + 3{\cal H}\Phi' + (2{\cal H}' + {\cal H}^2)\Phi &=& 
4\pi Ga^2\delta\bar p.
\end{eqnarray}
The bars in the perturbed fluid quantities indicate that we are using the gauge invariant expressions.

In UG we must perturb the equations,
\begin{eqnarray}
E_{\mu\nu} = 8\pi G\tau_{\mu\nu},
\end{eqnarray}
with the definitions,
\begin{eqnarray}
\label{ug1}
E_{\mu\nu} &=& R_{\mu\nu} - \frac{1}{4}g_{\mu\nu}R,\\
\label{ug2}
\tau_{\mu\nu} &=& T_{\mu\nu} - \frac{1}{4}g_{\mu\nu}T.
\end{eqnarray}

The perturbed equations of the UG equations (\ref{ug1},{\ref{ug2}) coupled to a fluid, using the gauge invariant formalism are:
\begin{eqnarray}
\label{pug1}
\Phi'' + 2({\cal H}' - {\cal H}^2)\Phi + \nabla^2 \Phi &=& 4\pi Ga^2(\delta\bar\rho + \delta \bar p),\\
\label{pug2}
\biggr(\Phi' + {\cal H}\Phi\biggl)_{,i} &=& - 4\pi G a^3\delta\bar u^i.
\end{eqnarray}
In obtaining these last expressions we have already used the fact that $\Phi = \Psi$. There are two important remarks on the equations (\ref{pug1},\ref{pug2}). First, there are two equations for three functions, $\Phi$, $\delta\tilde\rho = \delta\bar\rho + \delta\bar p$, and $\delta\bar u^i$. We will comment more on this issue later. The second important remarks is the the term $\nabla^2\Phi$ appears with the "wrong" sign compared with the GR case. We will also discuss more this fact later.

\subsection{Vacuum case: perturbations}

For vacuum, $\delta\bar\rho$, $\delta\bar p$ and $\delta\bar u^i$ are absent. The GR perturbed equations become,
\begin{eqnarray}
\label{ge1-tri}
- 3{\cal H}(\Phi' + {\cal H}\Phi) + \nabla^2\Phi &=& 0,\\
\label{ge2-tri}
\biggl\{\Phi' + {\cal H}\Phi\biggl\}_{,i} &=& 0,\\
\label{ge3-tri}
\Phi'' + 3{\cal H}\Phi' + (2{\cal H}' + {\cal H}^2)\Phi &=& 0.
\end{eqnarray}
On the other hand, the corresponding equations for UG are,
\begin{eqnarray}
\label{pug1-bis}
\Phi'' + 2({\cal H}' - {\cal H}^2)\Phi + \nabla^2 \Phi &=& 0,\\
\label{pug2-bis}
\biggr(\Phi' + {\cal H}\Phi\biggl)_{,i} &=& 0.
\end{eqnarray}

Let us first consider the de Sitter solution, for which, in the conformal time, $a \propto \frac{1}{\eta}$.
Equations (\ref{ge2-tri}) and (\ref{pug2-bis}) are the same, and it is satisfied in two cases: either $\Phi \propto 1/a$ or the perturbed quantities are spatial independent. Both hypothesis are consistent with each other. Hence, in both GR and UG cases, the solution of the perturbed equations are,
\begin{eqnarray}
\Phi = \frac{\Phi_0}{a},
\end{eqnarray}
$\Phi_0$ being a constant. 
The metric perturbation decreases as the universe expands, in agreement with the structure of the de Sitter space-time.

If now the Minkowski vacuum solution is inserted in the perturbed equations, $\Phi'' = 0$, leading to $\Phi \propto \eta\, +$ constant, both for GR and UG. Since the conformal time is, for the Minkowski case, equivalent to the cosmic time, the solution represents a growing mode.

\subsection{Introducing matter fields}

When matter is present many new features appear. First of all, many aspects of the problem depend if the conservation of the energy-momentum tensor is imposed or not. If the energy-momentum tensor conserves as in GR one of the first consequence is that the synchronous coordinate condition can not be use. The reason is the following. The unimodular constraint implies,
\begin{eqnarray}
h^\rho_\rho = 0.
\end{eqnarray}
If the synchronous coordinate condition $h_{\mu0} = 0$ is imposed, the unimodular constraint reduces to $h_{kk} = 0$ (a sum on the indice $k$ is understood). 
Using the conservation of the energy-momentum tensor, the UG equations reduce to the GR in presence of a cosmological constant. The perturbation of the field equations lead to the perturbed equation \cite{wei},
\begin{eqnarray}
\ddot {\tilde h} + 2H\dot {\tilde h} = 8\pi G\delta\rho,
\end{eqnarray}
with $\tilde h = h_{kk}/a^2$. If $h_{kk} = 0$ then $\delta\rho = 0$ and no perturbation is present.

Due to this property, a possibility is to use the gauge invariant formalism. This has been done in Ref. \cite{cai}. There, they found essentially the same equations of GR but with a new ingredient, a relation between the perturbed quantities due to the unimodular constraint. Hence, at perturbative level, even imposing the conservation of the energy-momentum tensor, UG has some distinguishing features.

If the conservation of the energy-momentum tensor is not imposed, the situation becomes more complex. The restriction to the use of the synchronous coordinate condition does not exist any more, but even so $h_{kk} = 0$. However, the density perturbation becomes connected to another metric perturbation $f = h_{ik,i,k}/a^2$. One important remark is that now there is no residual coordinate freedom associated to the synchronous coordinate condition. In fact, in GR the synchronous coordinate condition does not fix completely the coordinate system, and a residual, non-physical mode remains \cite{peebles}. This fact is reflected in the third order  (instead of a second order) differential equation for the density perturbation. However, in UG the unimodular condition eliminates this non-physical mode, and we end up with second order differential equations.

Of course, the gauge invariant formalism can always be used in UG, even with the modified conservation laws. However, there is a technical issue. As we can inspect from equations (\ref{pug1},\ref{pug2}), in the perturbed UG field equations there are two equations for three unknown functions; a new independent equation is needed. This new equation comes from the modified conservation law \cite{fabris1}. Using the gauge invariant formalism, to determine this new independent equation is a quite involved technical issue, while it is somehow direct using the synchronous coordinate condition. This has been done in Ref. \cite{fabris1}, where it was obtained the equation,
\begin{eqnarray}
\label{pe}
\ddot f + 3H\dot f - \frac{k^2}{3a^2}f = 0.
\end{eqnarray}
In this equation, $k$ is the wavenumber associated with the perturbation..

The final solution in terms of the conformal time ($\eta \propto t^{1/2}$) reads
\begin{eqnarray}
f = A\frac{\sinh \frac{k}{\sqrt{3}}\eta}{k\eta} + B\frac{\cosh \frac{k}{\sqrt{3}}\eta} {k \eta}.
\label{fsolution}
\end{eqnarray}
This solution reveals an exponential growth of the perturbations even if the background corresponds to the radiative phase. This is due to the "wrong" sign with the $k$-dependent term in (\ref{pe}) which is related with the Laplacian operator. We have already remarked that in the gauge invariant formalism such "wrong sign" of the Laplacian operator
also appears, see (\ref{pug1}), and a similar behavior can be expected. We are currently analysing this issue.

In Ref. \cite{fabris1} it has been also shown that a possible viable cosmological model can be obtained in UG when the modified energy-momentum tensor conservation is retained. This model must be refined in many ways, but in general lines, the age of the  universe, the CMB radiation, the present accelerated phase, and the origin of the structures resulting from the gravitational collapse out of a homogeneous and isotropic universe are well predicted by this model.

\section{An extension of Unimodular Gravity: including scalar fields}

The most direct extension of GR is by including scalar fields. It can be a self-interacting field representing the matter sector. In this case, the we modify only the right hand side of the field equations. However, it can also be implemented by a non-trivial coupling with the geometric sector and, in this case, the implications are more profound.
A paradigmatic example is the Brans-Dicke (BD) theory,  whose field equations, in presence of a cosmological constant, is given by\cite{bd},
\begin{eqnarray}
R_{\mu\nu} - \frac{1}{2}g_{\mu\nu}R &=& \frac{8\pi }{\phi}T_{\mu\nu} + \frac{\omega}{\phi^2}\biggr(\phi_{;\mu}\phi_{;\nu}
- \frac{1}{2}g_{\mu\nu}\phi_{;\rho}\phi^{;\rho}\biggl) + \frac{1}{\phi}\biggr(\phi_{;\mu\nu} - g_{\mu\nu}\Box\phi\biggl) + g_{\mu\nu}\Lambda,\\
\Box\phi &=& \frac{8\pi T}{3 + 2\omega} + \frac{4}{3 + 2\omega}\Lambda,\\
{T^{\mu\nu}}_{;\mu} &=& 0.
\end{eqnarray}
In these equations, $\omega$ is a free coupling parameter. GR is recovered when $\omega \rightarrow \infty$. The present estimations indicate a very high value for $\omega$ \cite{will}. Even though, BD remain an intensive object of studies, and it can be connected with many other fundamental theories, like string theories\cite{cordas}.

The unimodular version of the Brans-Dicke (UBD) theory has been proposed in Ref. \cite{fabris2}. The deduction of the field equations follows closely the RG case, introducing the unimodular constraint through Lagrangian multipliers. The final equations read, 
\begin{eqnarray}
\label{bde1-ff}
R_{\mu\nu} - \frac{1}{4}g_{\mu\nu}R &=& \frac{8\pi}{\phi}\biggr(T_{\mu\nu} - \frac{1}{4}g_{\mu\nu}T\biggl) + \frac{\omega}{\phi^2}(\phi_{;\mu}\phi_{;\nu} - \frac{1}{4}g_{\mu\nu}\phi_{;\rho}\phi^{;\rho})\nonumber\\
&+& \frac{1}{\phi}(\phi_{;\mu\nu} - \frac{1}{4}g_{\mu\nu}\Box\phi),\\
\label{bde2-ff}
\Box\phi &=& \frac{1}{2}\frac{\phi_{;\rho}\phi^{;\rho}}{\phi} - \frac{\phi}{2\omega}R,\\
\label{bde3-ff}
(\phi R)^{;\nu} &=& \omega\biggr(\frac{\phi_{;\rho}\phi^{;\rho}}{\phi}\biggl)^{;\nu} + 32\pi \biggr({T^{\mu\nu}}_{;\mu} - \frac{1}{4}T^{;\nu}\biggl) + 3(\Box\phi)^{;\nu}.
\end{eqnarray}
In this case, as in the GR one, the usual conservation of the energy-momentum tensor has not been imposed. If the usual conservation laws are introduced, the BD equations in presence of a cosmological constant are recovered.

The UBD has many new features in comparison with the traditional BD theory. We will comment just one of them. In Ref.\cite{plinio} an extensive perturbative analysis of cosmological models obtained from the BD theory was carried out.
The vacuum cosmological solutions in UBD coincide with the BD cosmological solutions in presence of a cosmological constant, as it happens with the corresponding solutions in GR. In Ref. \cite{plinio}, the vacuum solutions in presence of a cosmological term in the BD have been shown to be stable. However, in the UBD case the vacuum solution are unstable in the interval $- 1/2 < \omega < 3/2$. This is an important difference, pointing out that, even if the background UBD solutions can be mapped in the BD solutions in presence of a cosmological constant (something we could expect from the experience with GR and UG), the perturbative behavior is sensitively different. The inequivalence at perturbative level of BD and UBD theories is more evident than in the GR and UG case. This seems to be due to the presence of the scalar field non-minimally coupled to the gravity sector.

In UBD the scalar perturbations in the vacuum case, using the synchronous coordinate condition (which is now allowed), may be expressed through the master equation,
\begin{eqnarray}
\label{master2}
(3 - 2\omega)\delta\ddot\phi - \biggr[3(1 + 2\omega)H - 8\omega\frac{\dot\phi}{\phi}\biggl]\delta\dot\phi+ \biggr[12(\dot H + H^2) - 4\omega\frac{\dot\phi^2}{\phi^2}\biggl]\delta\phi + \frac{1 + 2\omega}{a^2}\nabla^2\delta\phi = 0.
\end{eqnarray}
The existence of instability in the UBD case, at least for some values of $\omega$, can be directly seen from (\ref{master2}) by inspecting the relative sign of the second derivative and the Laplacian terms of $\delta\phi$. In some sense, it is connected with the sound speed in the perturbations which, in (\ref{master2}), becomes imaginary for $- 1/2 < \omega < 3/2$.

\section{Conclusions}

Unimodular Gravity has been extensively discussed in the literature due to the possibility of giving new insights to some of the most important problems appearing in General Relativity. In particular, in UG the cosmological constant is somehow hidden in the general structure of the theory and, as far as the usual energy-momentum tensor conservation is imposed, it appears explicitly as an integration constant. 
This is generally considered as, at least, an alleviation of the cosmological constant problem that plagues GR. At quantum level also, it has been argued that UG can shed some light in the issues that appear in GR\cite{esp}.

The discussion of the possible equivalence between UG (without cosmological constant) and GR (in presence of a cosmological constant) is intensive, see Refs. \cite{wei-ug,ant, esp} for example. At the cosmological background level, the equivalence
seems to be clearly set out. But, the situation is less clear at the level of cosmological perturbations. In our point of view, the main aspect to be stressed is the invariance of UG to the more restricted transverse diffeomorphic transformation, while GR is invariant by the full diffeomorphism group.

We have discussed the background and perturbative issues in UG comparing them to GR (always with a cosmological constant). In vacuum, UG provides the same results as GR also at perturbative level, even if following a different path. In presence of matter, however, the situation is much more complex, depending first if the usual conservation laws are retained or not. If not, the configuration is clearly different \cite{fabris1}, as it could be expected, but even if the usual conservation laws are preserved, some new features appear.
The difference becomes stronger and more pronounced if a non-minimally coupled escalar field is introduced. In doing so, we can generalize the Brans-Dicke theory to the unimodular Brans-Dicke theory. An example of the this strong difference is the appearance of unstable modes in UBD which do not exist in the BD case.

Many aspects of the perturbative analysis presented here must be extended. One example is the use of the full gauge invariant formalism for the cases where matter is present and the conservation of the canonical energy-momentum tensor is not retained. But, the results exposed here seem to reinforce that the equivalence between UG and RG, even at classical level, is not complete, mainly when the perturbations are included.

\bigskip
\noindent
{\bf Acknowledgments:} We thank CNPq, FAPES and CAPES for partial financial support. We thank also Luiz Filipe Guimarães for his careful reading of the text.

\end{document}